# 3D reconstruction of the spatial distribution of dislocation loops using a triangulation approach

Hongbing Yu [1], Xiaoou Yi [2], Felix Hofmann [1]*

(1) Department of Engineering Science, University of Oxford, Parks Road, Oxford, OX1 3PJ, UK

(2) School of Materials Science and Engineering, University of Science and Technology Beijing, Beijing, 100083, China

**Abstract**

We propose a new approach for reconstructing the 3D spatial distribution of small dislocation loops (DLs) from 2D TEM micrographs. This method is demonstrated for small DLs in tungsten, formed by low-dose ion-implantation, that appear as circular spots in diffraction contrast images. To extract the 3D position of specific DLs, their 2D position in multiple weak-beam dark-field TEM micrographs, recorded at different tilt angles, is fitted. From this fit the geometric centre and size of each DL in each micrograph can be extracted. Using a forward prediction approach each specific DL is identified in all the 2D projections. A system of linear equations can then be setup, linking the 3D position of each DL to its 2D position in each projection. If more than 2 projections are available this system of equations is over-determined, and the 3D position of each DL is found by least-squares fitting. The results are in good agreement with the damage microstructure recovered using a generalized weighted back-projection method. Importantly the triangulation approach requires fewer projections and is less sensitive to the angular range covered by the projections. Our new method is also less sensitive to contrast variations due to local deviations from the diffraction condition. These advantages, as well as the accuracy of the triangulation method, are discussed in detail.

* Corresponding author: felix.hofmann@eng.ox.ac.uk

# 1 Introduction

Irradiation-induced nano-scale lattice defects play a key role in the degradation of mechanical and thermal transport properties of materials used in nuclear reactors [1,2]. The evolution of the size and density of these defects at low irradiation doses (before the overlap of displacement cascades) provides an opportunity to gain insight into the fundamental mechanisms governing the formation and evolution of irradiation-induced damage [3–5]. Transmission electron microscopy (TEM) is one of the most powerful tools to observe the formation and evolution of these few nanometer large defects [6,7]. For instance, in-situ TEM irradiation, has been used extensively to reveal the dynamics of lattice defects such as dislocation loops (DLs) and stacking fault tetrahedra (SFT) formed during the collapse of displacement cascades in the early stage of irradiation[3,5,8,9]. Weak-beam dark-field (WBDF) TEM techniques have been widely used to characterize the size, Burgers vector and nature of DLs and SFT at different dose levels [3,5,7,10–12]. They are generally considered to provide the most accurate size information for small irradiation-induced lattice defects [13]. However, TEM micrographs are only 2D projections of the 3D volume. The position information in the thickness direction is lost, severely hampering the analysis of 3D defect interactions.

Experimental observations [14] and defect simulation [15,16] have shown that elastic interactions between lattice defects play an important role in the microstructural evolution of ion-irradiation induced damage in tungsten. These elastic interactions depend on the relative 3D positions of defects, as well as their distance from the sample surface. Thus far, defect positions have been estimated solely by considering the separation of defects in 2D TEM micrographs, neglecting the depth information[14,16]. Furthermore, ion irradiation produces a non-uniform damage distribution in the depth direction. This effect should be accounted for when considering the evolution of microstructure with ion fluence, however, due to a lack of 3D spatial resolution, this has not been possible thus far [3,5,9]. Thus, to gain a more complete understanding of the formation and evolution of irradiation-induced damage, approaches that allow the determination of the 3D spatial distribution of defects are urgently needed.

Numerous tomography techniques have been proposed to recover 3D microstructure from a series of 2D projection images. Back-projection and simultaneous iterative reconstruction techniques (SIRT) are perhaps the most widely used for bright field tomography, particularly on biological samples [17,18]. For high quality reconstructions they require a large number of projections (more than 100) collected over a wide range of tilting angles (at least ±60°) with ~1° or smaller angular step size [19,20]. In TEM tomography high tilting angles are problematic as they increase the electron path length through the sample and cause shadowing effects when conventional TEM holders and samples are used [20]. Indeed, even if large tilt angles are reached, both back-projection and SIRT still suffer from "missing wedge" artefacts. In the last two decades, advanced reconstruction schemes such as discrete algebraic reconstruction technique[19,21] (DART) and total variation minimization [21–24] (TVM) method have been introduced. These methods effectively address the missing wedge problem and yield acceptable reconstructions, even for a reduced number of projections, by making use of prior knowledge. For the characterization of crystalline materials, and particularly lattice defects within them, the use of TEM techniques that rely on diffraction contrast is attractive. However, diffraction contrast from crystals generally does not satisfy the projection requirement and can cause substantial artefacts when conventional reconstruction approaches are used[20]. As a result, the application of 3D TEM tomography to imaging of crystal defects via their associated strain fields has only enjoyed limited success. For example, careful experiments, where the scattering vector (g) and deviation parameter ($s_g$) were held constant for a number of tilt angles, have been successfully used to image 3D structures of discrete dislocations from both WBDF or bright field (BF) images [25–28]. For very small defects, such as small DLs formed in low dose implanted metals, a further complication is that the contrast is very sensitive to subtle changes in deviation parameter, as well as changes of the depth of the DL in the TEM foil [4,29]. In pure W it has been reported that even a slight change in $s_g$ or loop depth can result in the transition of a small loop from visible to invisible, as well as alter loop size and morphology [5,10]. This means that sample tilt alone, or any local deviation from the desired diffraction conditions due to e.g. localized bending, sample quality or small rotations, can change the morphology, size and visibility of a DL. This experimental uncertainty means that the image of the same DL recorded at different tilt conditions will most certainly violate the projection principle. Furthermore, changes in $s_g$ and loop movement under electron illumination[3,5] can also cause imperfect alignment of projections, leading to blurring and an overestimation of defect size in 3D reconstructions.

Here we propose a triangulation approach to address this challenge. First, the image associated with each specific DL is identified in all WBDF micrographs, of the same sample volume, recorded at different tilts. A system of linear equations is then set up linking the 3D position of each DL to its 2D position in each projection. Since more than 2 projections are used, this system of equations is overdetermined and least-squares fitting is employed to find the 3D position of each DL. Considering self-ion damaged tungsten as an example, the results of the triangulation method are compared to the 3D damage microstructure recovered using weighted back-projection. This comparison reveals several advantages of the triangulation approach and allows an assessment of its robustness with regard to experimental uncertainties.

## 2 Materials and Method

## 2.1 Materials

TEM tomography data, i.e. 2D micrographs of DLs at different tilt angles, was recorded from a 150 keV self-ion implanted ultra-high purity W TEM foil (99.999% at. wt.). The sample was thinned to electron transparency by twin-jet electropolishing, using an electrolyte of 0.5 wt% NaOH aqueous solution close to 0 °C, followed by 3 rinses in methanol. In-situ irradiation was then performed with 150 keV self-ions at a temperature of 30 K to a fluence of $1.25 \times 10^{12} ion/cm^2$ on the IVEM-Tandem facility, Argonne National Laboratory. The damage dose corresponds to ~ 0.01 displacements per atom (dpa) at the peak of the damage profile according to SRIM calculation (quick K-P mode, displacement energy = 68 eV) [30]. At 30 K, the displacement cascades are effectively frozen. Data was collected in the as-irradiated condition and following a warm-up of the sample to room temperature. Here we use the data collected at room temperature. Using a double tilt holder, TEM micrographs were recorded under a (**g**, 5.25 ~ 6.25**g**) weak-beam dark-field condition with **g** = (110). The grain under consideration had approximately [001] surface normal, and the specimen was tilted around both α and β tilting axes from α = -36° and β = 20.2° to α = 32° and β = -19.6°. This is equivalent to tilting around the **g** vector from -41° to 37°. Care was taken to ensure that the **g** vector was maintained as (110) and the WBDF condition unchanged. This is important, as both the **g** vector and deviation parameter ($s_g$) will affect the visibility and apparent size of DLs. Overall, 19 WBDF TEM micrographs were recorded at different tilting conditions, as shown in supplementary figure S1, with a constant pixel size of 0.43nm by 0.43 nm.

## 2.2  3D triangulation approach

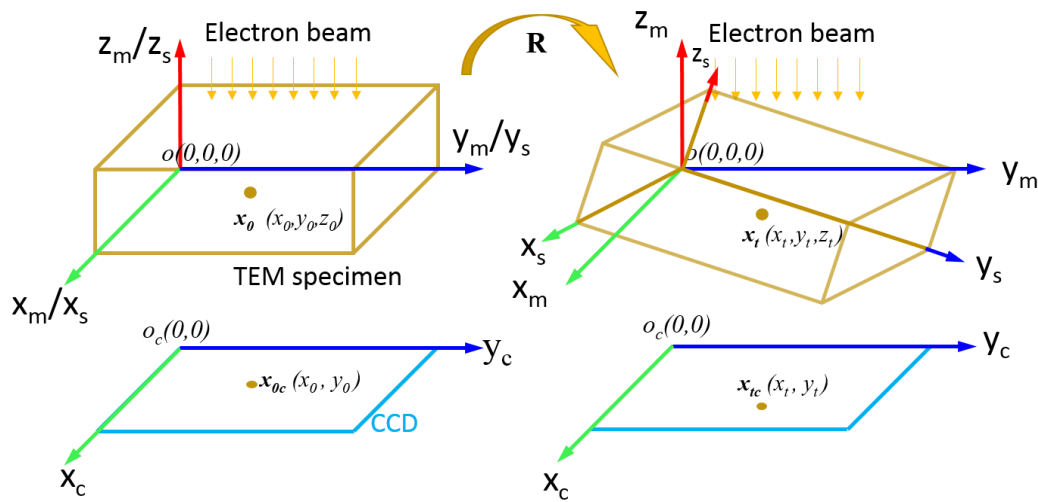

Fig. 1 Schematic diagram showing the geometric layout of electron beam, specimen, and CCD camera without and with tilting, and the definition of the coordinate systems of the specimen, microscope, and CCD camera.

To recover the 3D damage microstructure from the recorded 2D WBDF TEM micrographs, we propose the use of a triangulation approach. Fig. 1 shows a diagram illustrating the coordinate systems attached to the microscope, camera, and specimen and how sample tilting modifies the position at which features in the sample will appear on the detector. A right-handed coordinate convention is used throughout. The $x_m$- $y_m$- $z_m$ the coordinate system is attached to the microscope, and its origin is placed at the rotation centre. The CCD camera has coordinate system $x_c$- $y_c$ that corresponds to a projection of the microscope's $x_m$- $y_m$ coordinates along the $z_m$ direction. That is, if a feature has a position $\mathbf{x_0}$ ($x_0$, $y_0$, $z_0$) in microscope coordinates, its image on the CCD camera will be located at $\mathbf{x_{0c}}$ ($x_0$, $y_0$). The specimen's coordinate system ($x_s$- $y_s$- $z_s$) coincides with the microscope's system when no tilting is applied. In a TEM, the holder tilting axes generally do not coincide with the coordinate axes of the microscope or the CCD camera. Here we define that the α (major) tilting axis is a unit vector of $[t_1^\alpha \ t_2^\alpha \ 0]$ and the β tilting axis is a unit vector of $[t_1^\beta \ t_2^\beta \ 0]$, where both tilting axes are referenced to the microscope coordinate system for zero tilt. It should be noted that these two tilting vectors will vary from microscope to microscope. At the IVEM- Tandem facility, the α and β tilting axes respectively are [-0.90 0.44 0] and [0.44 0.90 0]. These tilts can be captured by a rotation matrix **R**:

$$\mathbf{R} = \begin{bmatrix} R_{11} & R_{12} & R_{13} \\ R_{21} & R_{22} & R_{23} \\ R_{31} & R_{32} & R_{33} \end{bmatrix} = \mathbf{R}_\alpha \mathbf{R}_\beta, \tag{1}$$

where $\mathbf{R}_\alpha$ and $\mathbf{R}_\beta$ are the rotation matrices for the α tilt and β tilt, respectively. The calculation of the matrix capturing rotation by a given angle about a unit vector is provided in Appendix 1.

After an arbitrary tilt, the coordinates, $\mathbf{x}_t$, in the microscope frame of a DL at position $\mathbf{x}_0$ in the sample are given by:

$$\mathbf{x}_t = \mathbf{R}\mathbf{x}_0 \tag{2}$$

And coordinates of the DL in the CCD camera's coordinate system are

$$\mathbf{x}_{tc} = \widetilde{\mathbf{R}}\mathbf{x}_0 \tag{3}$$

where $\widetilde{\mathbf{R}} = \begin{bmatrix} R_{11} & R_{12} & R_{13} \\ R_{21} & R_{22} & R_{23} \end{bmatrix}$.

If the coordinates of the DL, $\mathbf{x}_0$, are unknown, the position of the DL can be worked out from any two or more projections recorded at different, known tilting angles. If we have N (N≥2) projections available, the coordinates of the DL in the *n* th (n = 2, 3, …, N) projection are $\mathbf{x}_t^n$, and the rotation matrix for the *n* th (n = 2, 3, …, N) projection is $\mathbf{R}^n$, a system of equations can be established which can be written in the form of the matrix,

$$\mathbf{x}_{tc}^N = \mathbf{A}\mathbf{x}_0, \tag{4}$$

where

$$\mathbf{x}_{tc}{}^N = \begin{bmatrix} [\mathbf{x}_{tc}{}^1] \\ [\mathbf{x}_{tc}{}^2] \\ \vdots \\ [\mathbf{x}_{tc}{}^N] \end{bmatrix} \text{ and } \mathbf{A} = \begin{bmatrix} [\widetilde{\mathbf{R}}^1] \\ [\widetilde{\mathbf{R}}^2] \\ \vdots \\ [\widetilde{\mathbf{R}}^N] \end{bmatrix}.$$

From this, it is clear that at least two projections are needed to recover $\mathbf{x}_0$. If N >2 projections are acquired, the system of equations is over-determined. Here, least-squares fitting is used to find a solution for $\mathbf{x}_0$. This can be readily achieved using the pseudo-inverse of matrix $\mathbf{A}$ [31].

$$\mathbf{x}_0 = [(\mathbf{A}^t\mathbf{A})^{-1}\mathbf{A}^t] \, \mathbf{x}_{tc}{}^N, \tag{5}$$

where $\mathbf{A}^t$ is the transpose of $\mathbf{A}$.

## 2.3 Forward prediction

It is straightforward to calculate the depth of a specific DL once the position of this DL in two or more projections is known. However, identifying the image associated with a specific DL in multiple different projections is challenging. Image cross-correlation[32] and space invariant feature transformation[33,34] have been commonly used to match or recognize multiple instances of a specific feature[35]. However, there are many different DLs in each projection that have similar morphology, size, and intensity. Furthermore, the same DL will have slightly varying size, intensity, visibility and morphology when viewed in different projections as illustrated in supplementary figure S2. This complicates the use of digital image-correlation-based techniques for identifying the image of a specific DL in multiple projections. Instead, we propose a forward prediction approach to do the loop registration.

For any DL we can calculate the anticipated position in the camera coordinate frame if we know the 2D position of the DL in one projection and its depth in the sample coordinate frame. Since the depth of a given DL is not known a priori, we start with an initial guess of the depth. If the guess is close to the actual depth of the DL, the predicted position of the DL in other projections will match the position observed in the experimentally-recorded projections. However, if the guess is incorrect, the predicted 2D position for different projections will either not match the position of any DL or, wrongly, match the position of a different DL. Matching of the 2D position of a different DL will only occur sporadically. This means that the number of projections where a DL is found reaches a maximum when the guessed depth is close to actual value for that DL, as shown in Fig. 2. We note that a plateau rather than a single maximum point is found in the profile in Fig. 2. The best initial estimate of the loop depth is chosen as that with the smallest variance between predicted and measured 2D positions. Now each DL can be identified and labelled in all the projections. Then the system of linear equations described in Eq. (4) can be set up and solved using Eq. (5). Since the visibility of DLs may also change during tilting, even if **g** and the $s_g$ are kept constant, some DLs are only visible in a subset of projections. Here only loops that appear in 7 or more projections are considered.

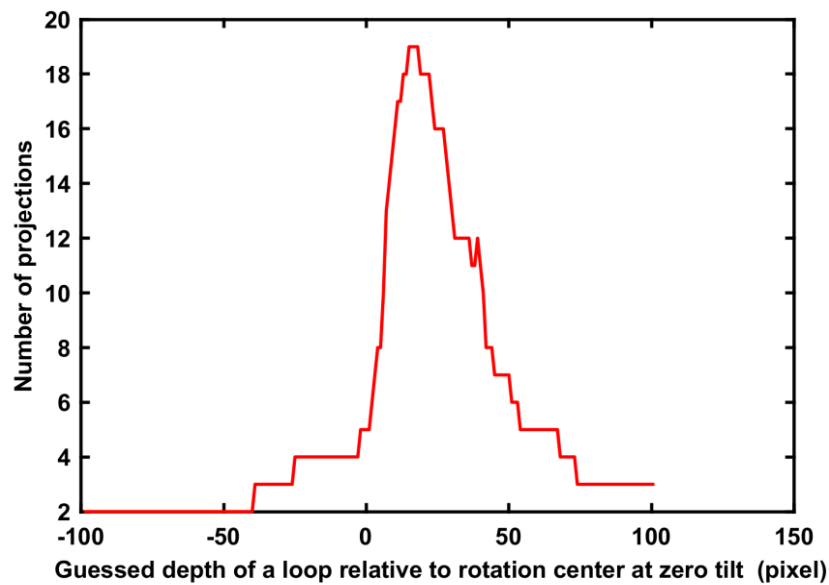

Fig. 2 The number of projections where a dislocation loop is matched versus the guess of depth of the dislocation relative to rotation centre at zero tilt.

## 2.4 Digital image processing

Before 3D reconstruction, the raw WBDF TEM micrographs were pre-processed using the approach of Mason et al [16] to remove the background signal. Variations in background intensity arise due to changes in foil thickness and residual stress-induced sample bending. They make automated DL detection challenging as e.g. bright background signal may be counted as foreground, while a dim DL spot on a dark background may be ignored, if a global threshold is applied without first removing the background signal. The approach of Mason et al can remove these background variations, while ensuring that the shape of foreground features is correctly retained (Fig 3 (b)). Finally a global threshold, determined using the Ridler & Calvard [36] thresholding algorithm, is applied to separate out the foreground features of interest.

Next, each DL was fitted with a 2D Gaussian function, as shown in Fig. 3 (c), to determine the 2D position and size of each DL. From this a binary image of the DL population was produced, using the half maximum of the peak intensity of the fitted Gaussian function as the adaptive threshold for each fitted DL (Fig. 3(d)).

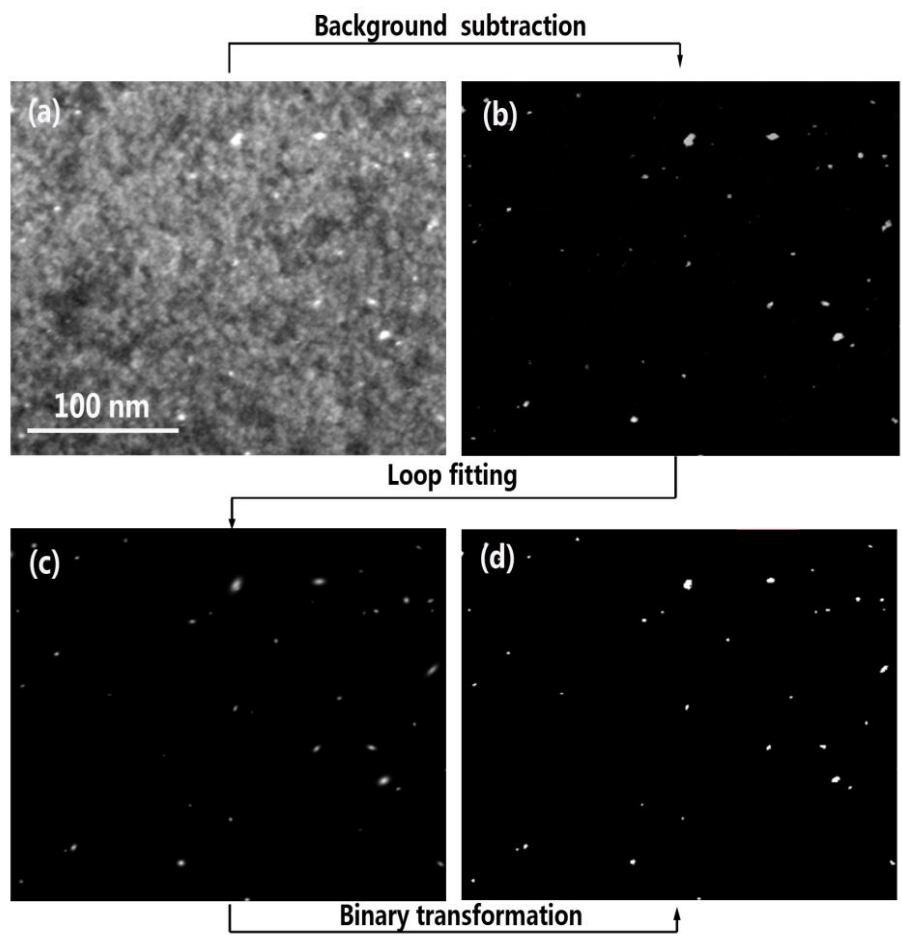

Fig. 3 Digital image processing. (a) Original image, (b) background subtracted, (c) dislocation loops with intensity higher than a certain threshold individually fitted with 2D Gaussian functions, (d) adaptive thresholding for each dislocation loop by half of the maximum intensity and convert the image to binary image.

The last step before the forward prediction is to align all the micrographs to a common rotation centre, here chosen as a DL prominently visible in all projections. Using image cross-correlation all micrographs were aligned such that the selected DL remained at the (0, 0) position. To reconstruct the 3D DL population, the forward prediction approach (section 2.3) was first used together with the binarized micrographs to perform the loop registration. Then, the accurate 2D position of each specific DL in each projection, determined using 2D Gaussian fitting, was used to determine the 3D position of each loop (section 2.2).

## 2.5 Back-projection

To provide a comparison for the results of the triangulation method, a 3D reconstruction using a generalized back-projection approach, was also considered. Back-projection refers to the process of projecting the intensity in a 2D projection back into the 3D volume along the projection path. By summing the back-projected intensity from multiple 2D projections, the 3D structure from which the 2D projections were acquired can be recovered. Fig. 4 (b) demonstrates, for a 2D example, the reconstruction by back-projection of the object in Fig. 4 (a), using a small number of projections (19) recorded at angles between -40° and 40°. Fig. 4 (b) clearly shows line-artefacts corresponding to the paths along which specific 2D projections were back-projected. These can be avoided by increasing the angular range covered by projections (ideally a full 180°) and reducing the angular interval between projections. A blurring of the reconstructed features is also evident, caused by a higher weighting of low-spatial frequencies during the back-projection process [37].

This issue can be tackled by two different methods. The first is filtered back-projection which involves applying a frequency filter to the recorded projections to reduce low frequency components before back-projection[37]. The second is weighted back-projection where the point spread function associated with the back-projection is removed from the reconstructed 2D or 3D images using a deconvolution[38]. The point spread function can be obtained by forward-projecting a delta function (Fig. 5 (a)) and then back-projecting it in the same way as the object to be reconstructed. Fig. 5 (b) shows the 2D point spread function for the back-projection shown in Fig. 4 (b). A number of different algorithms can be used for the 3D deconvolution[39]. In this paper, the Richard and Lucy iterative method was chosen. Fig.4 (c) and Fig. 4 (d) show reconstructions using filtered back-projection and weighted back-projection respectively of the image in Fig. 4 (a). Filtered back-projection substantially reduces the blurring of the reconstructed image, but the stripe artefacts, due to the small number of projections (19) and limited angular range, are still clearly visible. Weighted back-projection does a better job for a small number of projections. In this paper, weighted back-projection for dual-axis tilting was used to reconstruct the 3D spatial distribution of DLs for a comparison with the triangulation-based reconstruction approach.

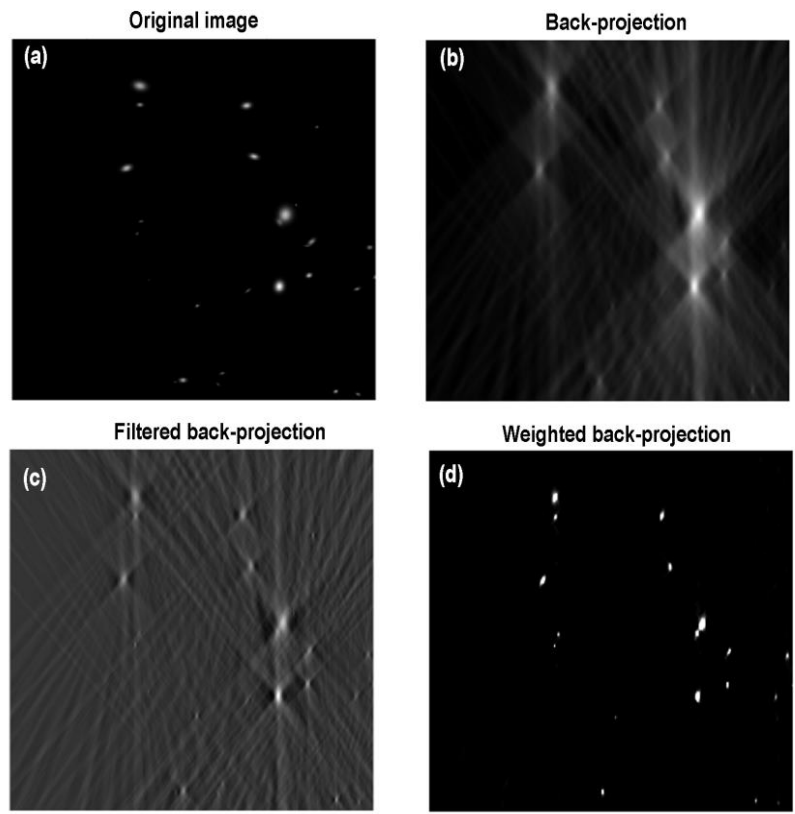

Fig. 4 A 2D case demonstrating back-projection reconstruction and two methods to improve the quality of back-projection. (a) Original image, (b) back-projection from 19 projections recorded at angles ranging from -40° to 40°, (c) filtered back-projection, (d) deconvolution of the point spread function as shown in Fig. 5(b) from the back-projection reconstructed image.

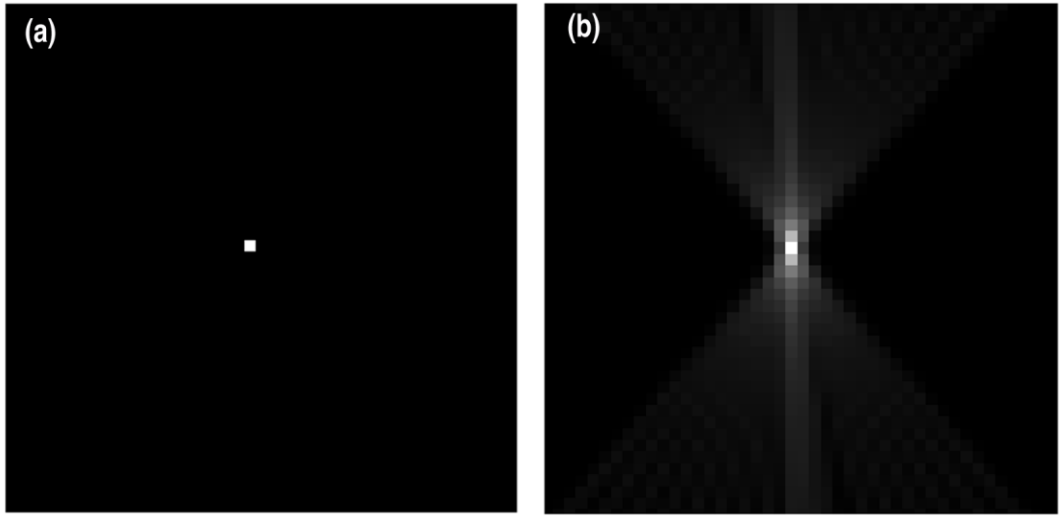

Fig. 5 (a) 2D delta function. (b) Reconstruction for the single point in (a) by back-projection from projections recorded at the same set of angles as in Fig. (4).

## 3   Results and discussions

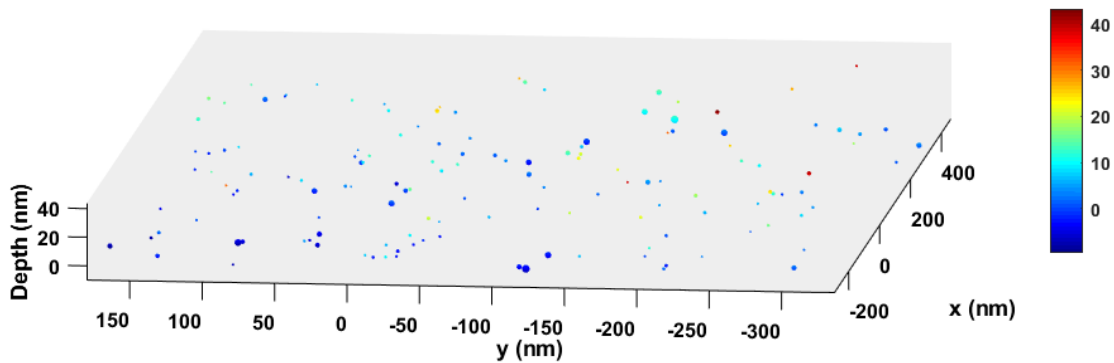

Fig. 6 3D spatial distribution of dislocation loops recovered by least-squares fitting approach. The colour map represents the depth relative to the rotation centre. The size of each marker represents dislocation loop size.

Fig. 6 shows the 3D spatial distribution of DLs recovered from a series of 2D micrographs, recorded at different tilt angles, using the triangulation approach. The spherical markers represent DLs in 3D space. Their diameter is proportional to the determined DL size and their color indicates their depth with respect to the rotation centre. An animated version of this plot is provided in supplementary movie 1. It can be clearly seen that DLs are confined to a shallow layer of ~ 30 nm thickness, consistent with the damage layer thickness predicted by SRIM for irradiation of tungsten with 150 keV self-ions.

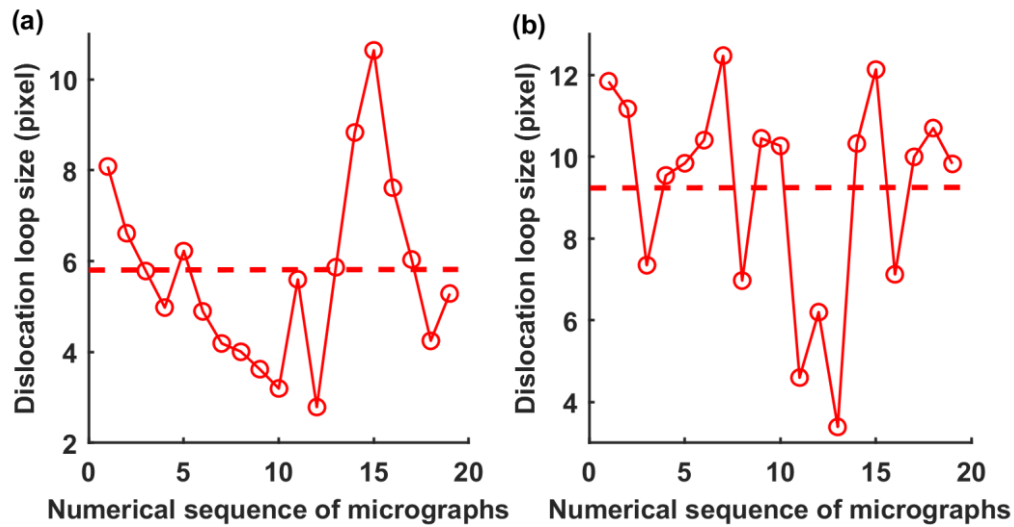

Fig. 7 Variation of the apparent size of two dislocation loops from projection to projection. The projections are ordered by ascending α tilting angle. Superimposed is a dashed line indicating the average size of each loop.

An important question concerns the determination of DL size. For small DLs, up to 10 nm in size, the width at half intensity along the direction of maximum extent of the defect image is commonly used as a measure of DL size [13]. Considering two specific DLs, Fig. 7 shows that their apparent size varies significantly from projection to projection. These changes merit further discussion. Under diffraction contrast conditions, changes in the apparent size of DLs can be divided into two categories: Geometric and experimental. Geometric variations refer to size changes caused by viewing of the DL strain field from different angles. In this case, a smooth, sinusoidal variation with tilting angle is expected. Experimental dependence refers to experimental uncertainties that affect the apparent size of a DL, such as localized bending, surface roughness, thickness contours, residual stresses, and change of the distance of a DL to sample surface along the elctron beam direction [7]. These experimental effects are difficult to quantify. The substantial variation of DL size from projection to projection in Fig. 7 suggests that experimental factors play an important role. To reduce the errors induced by experimental factors, the size of each DL we report is determined by averaging its apparent size over all projections where it is visible. These substantial variations in apparent DL size advise caution for studies that rely on single WBDF images of a defect microstructure to determine DL size distribution.

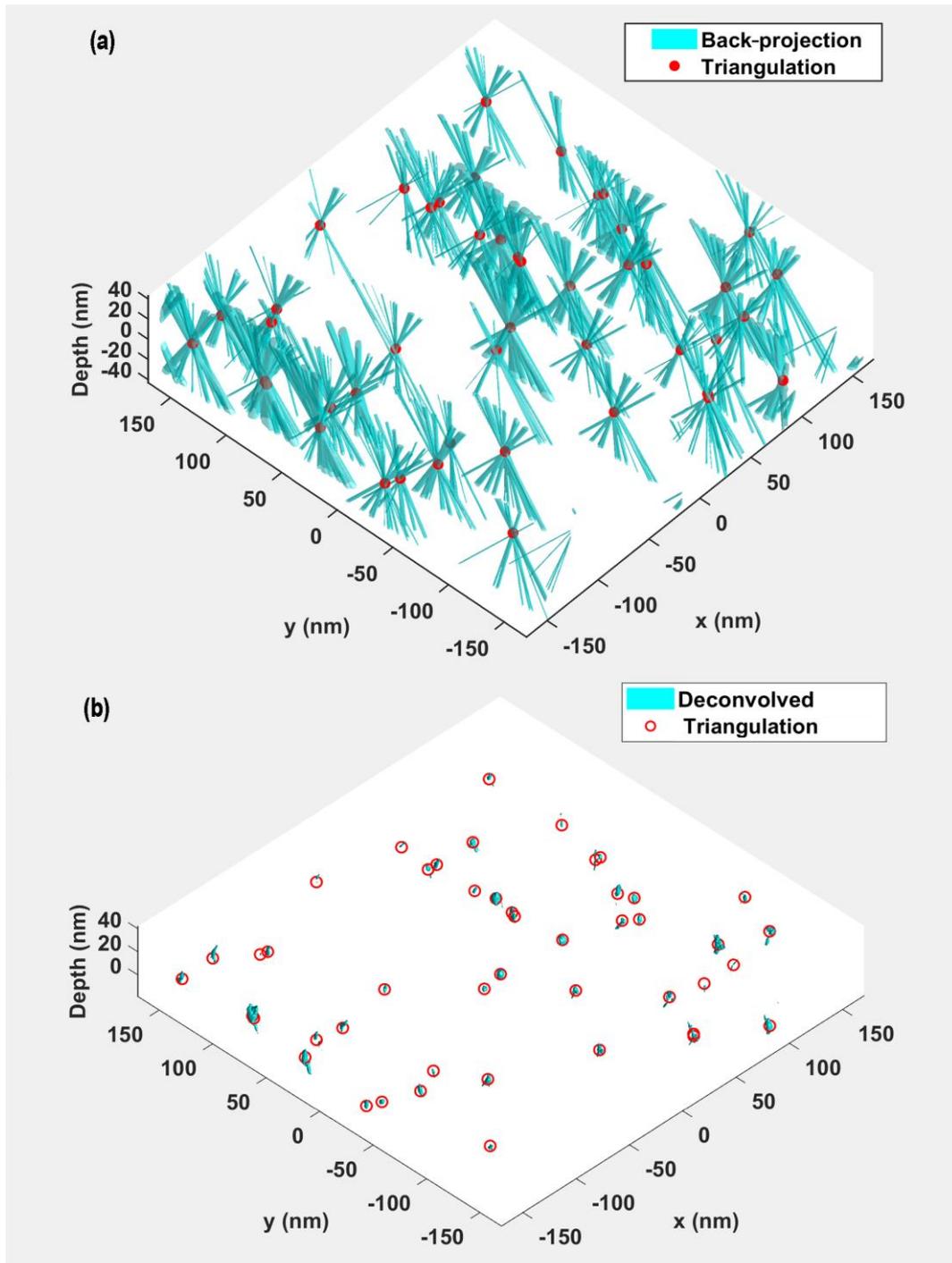

Fig. 8 Superposition of the 3D reconstructions by least-squares fitting method and back-projection. (a) Back-projection only, (b) deconvolve a 3D point spread function from the back-projected 3D volume.

Fig. 8 shows a comparison of the spatial distribution of DLs recovered using the triangulation approach and the back-projection scheme (see supplementary movies 2 and 3 for 3D animations). Fig. 8 (a) demonstrates the connection between the triangulation approach and the back-projection. The line-artefacts (visible in turquise), represent the paths along which the images of a specific DLs were back-projected into the 3D volume. These paths can be represented by a system of linear equations. The triangulation approach finds the least-squares solution to this system of linear equations. Fig. 8 (b) shows the comparison of DL positions reconstructed using the triangulation approach and weighted back-projection. The 3D spatial distribution of DLs clearly agrees very well.

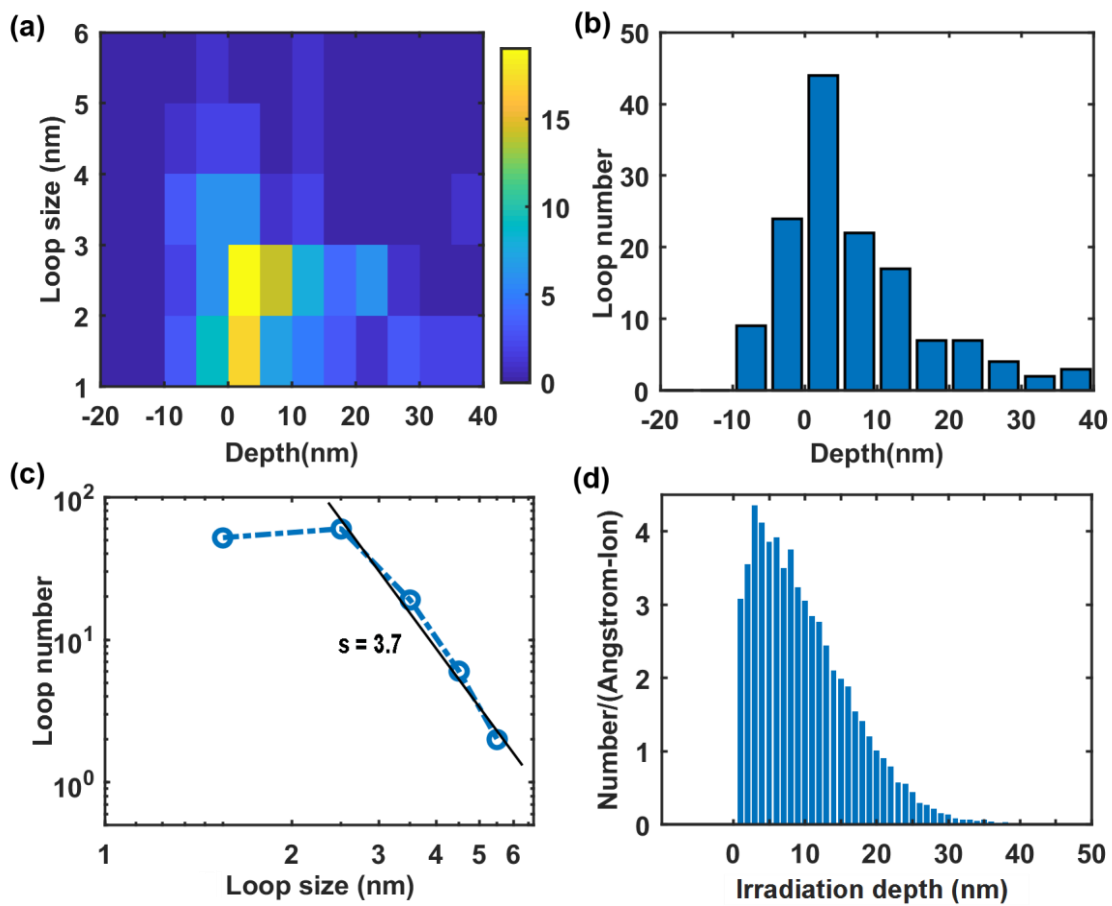

Fig. 9 (a) Number of dislocation loops in the reconstructed volume as a function of depth and loop size. (b) and (c) are histograms obtained by summing the 2D plot of (a) along the dimension of loop size and depth, respectively. The size distribution in (c) is plotted on a log-log scale. (d) Damage profile for 150 keV self-ion implantation of pure W predicted by SRIM.

Now that 3D DL positions and sizes are known, this data can be analysed in more detail. Fig. 9 (a) shows the number of DLs in the reconstructed volume, plotted as a function of loop depth (in 5 nm increments) and loop size (1 nm increments). DLs are distributed over a 50 nm thick region in the depth direction, ranging from -10 nm to 40 nm relative to the rotation centre, where the sample surface is at a depth of approximately -10 nm. DL sizes range from 1 nm to 6 nm. Interestingly the size distribution of DLs varies as a function of depth in the sample. Furthermore, the depth profile of DL density varies depending on the loop size. This suggests that the depth-dependent variation of defect population is not a simple function of damage dose, information that is lost when only considering 2D WBDF micrographs recorded at a single tilt.

By integrating the 2D data in Fig. 9 (a) either in the depth or in the loop size directions, histograms of depth dependence of DL population or size distribution can be obtained (9 (b) and (c) respectively). Fig. 9 (d) shows the damage profile for 150 keV self-ion implantation in W predicted using the SRIM code. Comparison of Fig. 9 (b) and (d) shows that the anticipated damage profile matches the experimentally observed depth-dependence of DL number density quite well, except within 10 nm of the free surface. This suggests that the sample surface acts as a strong sink for small defects, an effect that has previously been observed in Kinetic Monte Carlo simulations of damage evolution in tungsten[15]. The size distribution of DLs, plotted on a log-log scale (Fig. 9(c)) follows a power law $f \sim A/d^s$, where f is the frequency of occurence, A is a constant, and d is the diameter of DL, s is the exponent. The exponent is 3.7, in good agreement with predictions from MD simulations [40]. For loops smaller than 2 nm a deviation from this power law is observed, in good agreement with previous observations from 2D micrographs [14]. This may be due to a lack of sensitivity of WBDF TEM measurements to these very small defects, which has been previously pointed out [29].

Next we consider the robustness of the triangulation method to the number of recorded projections. In principle, only two projections are required, however a greater number of projections reduces experimental uncertainty and aids loop registration. It is interesting to explore how many projections are needed to obtain a reliable reconstruction. Fig. 10 shows the depth of two DLs calculated from all the possible combinations of different numbers of projections picked from the measured 19 projections. The x-axis shows the number of projections used and y-axis the loop depth with error bars indicating the variance. For the DL in Fig. 10 (a) the variance is large when only two projections are selected, but reduces rapidly once more projections are used. A similar trend is observed for the DL in Fig. 10 (b). The inset images in Fig 10 (a) and (b) show a cross section of the corresponding loops in the 3D volume reconstructed by back-projection. It can be seen that variance is due to the paths, along which the images of the DL were back-projected into the 3D volume, not meeting at exactly the same position. This effect is particularly prominent in Fig. 10 (a). There may be several reasons for this misalignement, such as local variations of the deviation parmeter due to bending of the foil or loop hopping under electron beam illumination, which has been observed in pure Fe[3] and W[5]. As a result, the minimum number of projections required is strongly dependent on the experimental uncertainties and DL stability under the electron beam. From our measurements we estimate that 10 projections should be sufficient in most cases.

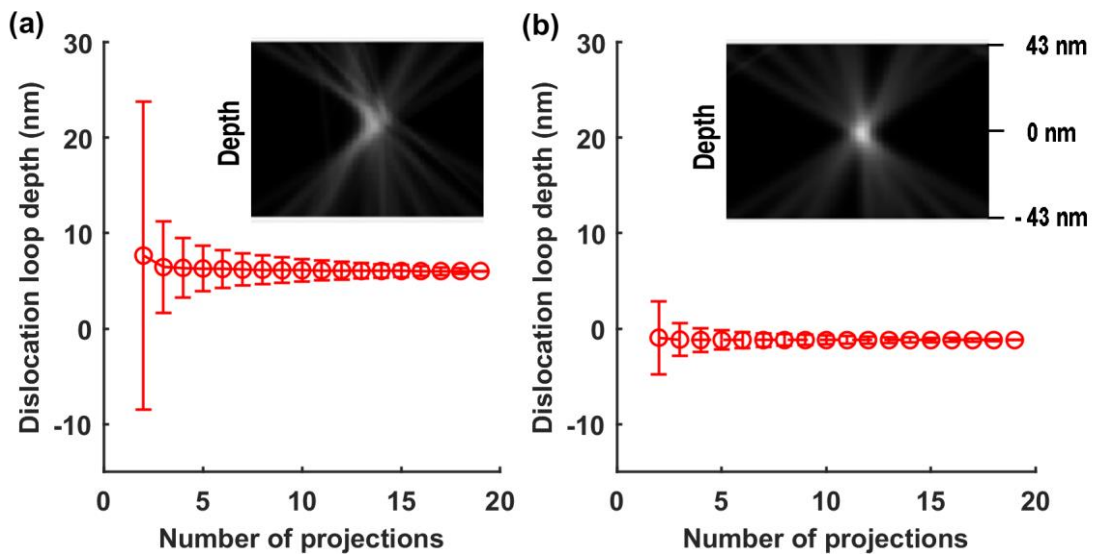

Fig. 10 The depths of two dislocation loops calculated from all the possible combinations of different numbers of projections out of the total 19 recorded projections. The circles mark the average depth calculated from all the possible combinations and the errors bars represent the variance. Inset images show cross-sections of the corresponding loops extracted from the back-projection reconstruction.

Misalignment also leads to a blurring of the DL images in back-projection reconstrutions, as seen in Fig. 10 (a). This in turn may lead to an over-estimation of loop size when back-projection approaches are used. The triangulation approach, where loop size is determined based on the apparent size of dislocation loops in 2D projections, avoids this effect. This benefit was also noted in a recent study where use of a triangulation approach reduced the reconstruction errors for line dislocations [41].

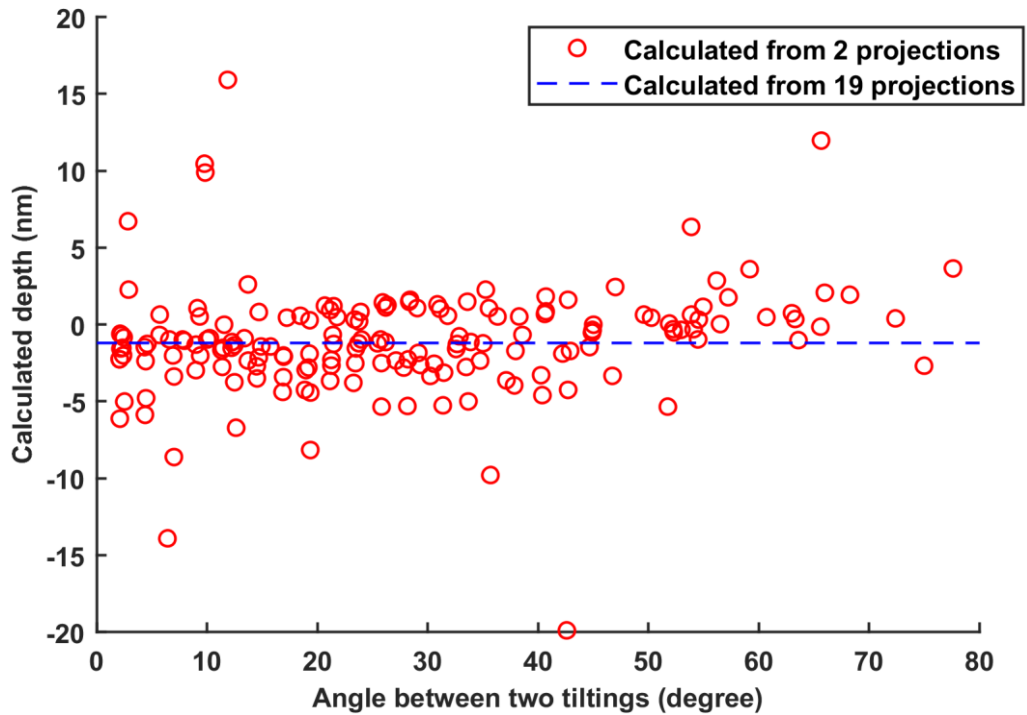

Fig. 11 The estimated depth of a specific DL versus the angle between the two projections used to estimate the depth of the DL. Superimposed is a dashed line indicating the depth determined by fitting of all 19 projections.

The angular range over which projections are collected is also expected to play an important role, with a small tilting range reducing the accuracy of depth deternination. Fig. 11 shows a plot of the depth of a DL, estimated from two projections, plotted as a function of tilting angle change between the two projections. Several outliers are present at angles below 15 degrees. A larger angles, above 50 degrees, an increase in scatter is also evident. This is mainly due to localized bending, which makes it difficult to maintain exactly the same diffraction condition when two projections are recorded at very different tilting angles. However, overall, the estimated depth shows surprisingly little dependence on the angle between the two projections. Thus, since the high angle tilting required for conventional reconstruction scheme is not needed for the triangulation approach, an angular tilt range of 30 to 40 degrees provides a good compromise.

**Conclusions**

We have developed a triangulation approach for determining the 3D distribution of small dislocation loops based on weak-beam dark field TEM micrographs recorded at a number of different tilt angles. As a demonstration this approach is applied to the study of the 3D damage microstructure in low-dose self-ion implanted tungsten. Using a forward prediction approach, each specific DL is identified in all the projections. A system of linear equations is then setup linking the 3D position of each DL to its 2D position in each projection. Least-squares optimisation is then used to find a solution for the 3D loop position. Using this 3D information the spatial correlation between defects, as well as size and depth-dependence of the dislocation loop population can be examined. Our analysis shows that the triangulation approach is more tolerant to contrast variations than back-projection approaches, and requires only a small number of projections that span a modest tilting angle range.

**Acknowledgements**

We thank Dr. Daniel Mason for his help in image processing. We thank Marquis Kirk, Pete Baldo and Edward Ryan for their help with the irradiations. We also thank Dr. Meimie Li for helpful discussion. We acknowledge funding from the European Research Council (ERC) under the European Union's Horizon 2020 research and innovation programme (grant agreement No 714697). The experiments of in situ ion irradiations at liquid helium temperature were carried out at Argonne National Laboratory (ANL), using the IVEM-Tandem Facility. These experiments in ANL were supported by a U.S. Department of Energy Facility funded by the DOE Office of Nuclear Energy, operated under Contract No. DEAC02-06CH11357 by U. Chicago Argonne, LLC.

**Appendix**

Given a unit vector $\mathbf{r}=(r_1, r_2, r_3)$, where $r_1^2+r_2^2+r_3^2=1$, the transformation matrix for a rotation by an angle of θ about $\mathbf{r}$, following a right-handed convention, is given by

$$R=\begin{bmatrix} r_1^2(1-\cos\theta)+\cos\theta & r_1r_2(1-\cos\theta)-r_3\sin\theta & r_1r_3(1-\cos\theta)+r_2\sin\theta \\ r_2r_1(1-\cos\theta)+r_3\sin\theta & r_2^2(1-\cos\theta)+\cos\theta & r_2r_3(1-\cos\theta)-r_1\sin\theta \\ r_3r_1(1-\cos\theta)-r_2\sin\theta & r_3r_2(1-\cos\theta)+r_1\sin\theta & r_3^2(1-\cos\theta)+\cos\theta \end{bmatrix}$$